\documentclass[a4paper]{scrartcl}

\usepackage{cite}
\usepackage[english]{babel}
\usepackage[utf8]{inputenc}
\usepackage[T1]{fontenc}
\usepackage{lmodern}
\usepackage{amsmath, amssymb}
\usepackage{amsthm}
\usepackage[fixlanguage]{babelbib}
\selectbiblanguage{english}
\usepackage[pdftex]{graphicx,color}
\usepackage{float}
\usepackage{caption}
\usepackage{subcaption}
\usepackage[dvipsnames,svgnames]{xcolor} 
\usepackage{tikz,pgf}
\usetikzlibrary{calc}
\usepackage[colorlinks=true,linkcolor=RoyalBlue,citecolor=PineGreen,urlcolor=blue]{hyperref}

\usepackage{booktabs}
\usepackage{fancyvrb}
\usepackage{xstring}

\newtheorem{theorem}{Theorem}
\newtheorem{lemma}{Lemma}
\theoremstyle{definition}\newtheorem{definition}[theorem]{Definition}

\usepackage{algorithm,algorithmicx,algpseudocode}

\algnewcommand\And{\textbf{and} }

\DefineVerbatimEnvironment{IsageV}{Verbatim}{gobble=2}
\DefineVerbatimEnvironment{Osage}{Verbatim}{gobble=2,formatcom=\color{black!50!white},fontsize=\scriptsize}
\newenvironment{Isage}{%
 \renewcommand{\FancyVerbFormatLine}[1]{\IfBeginWith{##1}{.}{....:  \noexpandarg\StrGobbleLeft{##1}{1}}{sage: ##1}}%
 \VerbatimEnvironment\begin{IsageV}}{\end{IsageV}}

\newenvironment{IOsage}{%
 \renewcommand{\FancyVerbFormatLine}[1]{\IfBeginWith{##1}{.}{....:  \noexpandarg\StrGobbleLeft{##1}{1}}%
 			{\IfBeginWith{##1}{*}{\textcolor{black!50!white}{\scriptsize{}\noexpandarg\StrGobbleLeft{##1}{1}}}{sage: ##1}}}%
 \VerbatimEnvironment\begin{IsageV}}{\end{IsageV}}

\newcommand{\graphicalOutput}[2][0.8]{\includegraphics[width=#1\textwidth]{#2}\newline}

\colorlet{col1}{LimeGreen}
\colorlet{col2}{Orchid}
\colorlet{col3}{Orange}
\colorlet{col4}{Cerulean}
\colorlet{col5}{Goldenrod}
\colorlet{col6}{LightGray}
\definecolor{colR}{rgb}{.932,.172,.172} 
\definecolor{colB}{rgb}{.255,.41,.884} 

\tikzstyle{vertex}=[circle, draw, fill=black, inner sep=0pt, minimum size=4pt]
\tikzstyle{smallvertex}=[circle, draw, fill=black, inner sep=0pt, minimum size=2pt]
\tikzstyle{edge}=[line width=1.5pt,black!50!white]
\tikzstyle{gridl}=[black!50!white]
\tikzstyle{axes}=[gridl,-latex]

\tikzstyle{lnode}=[circle,white,draw=black!60!white,fill=black!60!white,inner sep=1pt]
\tikzstyle{lnodesmall}=[circle,white,draw=black!60!white,fill=black!60!white,inner sep=1pt, font=\scriptsize]
\tikzstyle{redge}=[edge,colR]
\tikzstyle{bedge}=[edge,colB]

\colorlet{colvR}{black!70!white}
\tikzstyle{lnodeR}=[circle,colvR,draw=colvR,fill=white,inner sep=1.2pt, font=\scriptsize]
\tikzstyle{vertexR}=[circle,draw=colvR,fill=white,inner sep=0pt, minimum size=4.5pt]

\tikzstyle{arbitrarygraph}=[dashed,black!70!white,thick]
\tikzstyle{labelsty}=[font=\scriptsize]

\newcommand{\sage}{\textsc{SageMath}}
\newcommand{\flexrilog}{\textsc{FlexRiLoG}}

\newcommand{\blue}{\text{blue}}
\newcommand{\red}{\text{red}}

\DeclareMathOperator{\Upairs}{U}
\newcommand{\upairs}[1]{\Upairs(#1)}

\DeclareMathOperator{\CDC}{CDC}
\newcommand{\cdc}[1]{\CDC(#1)}

\newcommand{\RR}{\mathbb{R}}

\title{FlexRiLoG --- A SageMath Package for Motions of Graphs}
\author{Georg Grasegger\thanks{Johann Radon Institute for Computational and Applied Mathematics (RICAM), Austrian Academy of Sciences}
\and Jan Legerský\thanks{Johannes Kepler University Linz, Research Institute for Symbolic Computation (RISC)}
\thanks{Department of Applied Mathematics, Faculty of Information Technology, Czech Technical University in Prague}}
\date{}

\begin{document}
\maketitle
\begin{abstract}
	In this paper we present the \sage\ package \flexrilog\ (short for flexible and rigid labelings of graphs).
	Based on recent results the software generates motions of graphs
	using special edge colorings.
	The package computes and illustrates the colorings and the motions.
	We present the structure and usage of the package.
\end{abstract}

\section{Introduction}
A graph with a placement of its vertices in the plane is considered to be flexible if 
the placement can be
continuously deformed by an edge length preserving motion into a non-congruent
placement.
The study of such graphs and their motions has a long history 
(see for instance \cite{Burmester1893,Dixon,Kempe1877,Stachel,WalterHusty,Wunderlich1954,Wunderlich1976,Wunderlich1981}).
Recently we provided a series of results \cite{flexibleLabelings,movableGraphs} 
with a deeper analysis of the existence of flexible placements.
This is done via special edge colorings called
NAC-colorings (“No Almost Cycles”, see \cite{flexibleLabelings}).
These colorings classify the existence of
a flexible placement in the plane and give a construction of the motion.

\textbf{Basic definitions.}
We briefly give a more precise definition of flexibility of a graph. 
A \emph{framework} is a pair $(G,p)$
where $G=(V_G,E_G)$ is a graph and $p : V_G \rightarrow \RR^2$
is a \emph{placement} of $G$ in $\RR^2$.
The placement might be possibly non-injective but for all edges $uv \in E_G$ we require $p(u) \neq p(v)$. 

Two frameworks $(G,p)$ and $(G,q)$ are \emph{equivalent} if 
for all $uv \in E_G$,
\begin{align}\label{eq:equivalence}
	\| p(u) - p(v) \| = \| q(u) - q(v)\|\,.
\end{align}
Two placements $p,q$ of $G$ are said to be \emph{congruent} if \eqref{eq:equivalence} holds for all pairs of vertices $u,v \in V_G$.
Equivalently, $p$ and $q$ are congruent if there exists a Euclidean isometry~$M$ of $\RR^2$ such that $M q(v) = p(v)$ for all $v \in V_G$.

A \emph{flex} of the framework $(G,p)$ is a continuous path $t \mapsto p_t$, $t \in [0,1]$,
in the space of placements of $G$ such that $p_0= p$ and each $(G,p_t)$ is equivalent to~$(G,p)$.
The flex is called trivial if $p_t$ is congruent to $p$ for all $t \in [0,1]$.

We define a framework to be \emph{flexible} if there is a non-trivial flex in $\RR^2$.
Otherwise is is called \emph{rigid}.
We say that a labeling $\lambda: E_G\rightarrow \RR_{>0}$ of a graph~$G$ is \emph{flexible}
if there is a flexible framework $(G,p)$ such that 
$p$ induces $\lambda$, namely, $\|p(u) - p(v) \| = \lambda(uv)$ for all~$uv\in E_G$.
On the other hand,  $\lambda$ is \emph{rigid} if $(G,p)$ is rigid for all placements $p$ inducing~$\lambda$.
A flexible labeling $\lambda$ of a graph is \emph{proper} if there exists a framework $(G,p)$ such that 
$p$ induces $\lambda$ and it has a non-trivial flex with all but finitely many placements being injective. 
We call a graph \emph{movable} if it has a proper flexible labeling.

\textbf{Outline of the paper.}
We have given the necessary definitions.
Section~\ref{sec:package} describes the main functionality of the \flexrilog\ dealing with colorings and motions.
In this paper we do not provide the algorithms themselves but refer to the respective theorems and literature.
In Section~\ref{sec:movable} we describe how to use the package to ask for movable graphs.

\section{The Package}\label{sec:package}
\flexrilog\ \cite{flexrilog}
is a package for \sage\ running in versions 8.9 and 9.0 \cite{sagemath}.
The latest release of the package can be installed by executing:
\begin{Verbatim}
sage -pip install --upgrade flexrilog
\end{Verbatim}
The development version of \flexrilog\ can be found in the repository \cite{flexrilogGitHub},
where also other options of installation are described.

A convenient way of using the package instead of the \verb=sage= console is a Jupyter notebook
(coming with~\sage, launch by \verb=sage -n jupyter=).
The Jupyter notebook \verb=examples/flexrilog_Motions_of_Graphs.ipynb= in~\cite{flexrilogGitHub}
provides an interactive version of this paper. 

The package allows to check whether a graph has a NAC-coloring, in particular to list all of them.
A motion or flex obtained from a NAC-coloring can be constructed and displayed.
Moreover, it implements the results of~\cite{movableGraphs} regarding the existence of proper flexible labelings,
namely, the check of a necessary condition and construction of a proper flex from a pair of NAC-colorings.
There is also functionality providing tools for classification of all proper flexible labeling,
which is out of the scope of this paper (see \cite{ClassificationPaper} for details).

\subsection{Data Types}
The package provides data types in different classes for dealing with graphs, colorings and motions.
In order to use the data types provided by the package, they have to be loaded.
\begin{Isage}
  from flexrilog import FlexRiGraph, GraphMotion
\end{Isage}
The main object will always be of type \verb=FlexRiGraph=.
This class inherits properties of the standard \verb=Graph= from \sage\ 
and adds specific properties for investigations of flexibility and rigidity.
In this paper we focus on the flexibility part.
A \verb=FlexRiGraph= can be constructed by the following command.
\begin{IOsage}
  FlexRiGraph([[0,1],[1,2],[0,2]])
  *FlexRiGraph with the vertices [0, 1, 2] and edges [(0, 1), (0, 2), (1, 2)]
\end{IOsage}
Further constructions can be made via integer encoding described in \cite{ZenodoAlg} and via objects of the standard \verb=Graph= class.
\begin{IOsage}
  FlexRiGraph(graphs.CompleteBipartiteGraph(2,3))
  *Complete bipartite graph of order 2+3: FlexRiGraph with the vertices [0, 1, 2, 3, 4]
  *and edges [(0, 2), (0, 3), (0, 4), (1, 2), (1, 3), (1, 4)]
\end{IOsage}

Besides the class for graphs there is a class for colorings (\verb=NACcoloring=).
We do not discuss the class itself here but rather show how to compute colorings of graphs (see Section~\ref{sec:colorings}).
Furthermore, motions are stored in a third class, \verb=GraphMotion=. They are discussed in Section~\ref{sec:motions}.
The \verb=GraphGenerator= class stores the code for some important graphs from the area of rigidity and flexibility theory.
We do not go into detail but some of the graphs are used in the paper.

\subsection{NAC-Colorings}\label{sec:colorings}
NAC-colorings are a special type of edge colorings using two colors.
Unlike proper edge colorings in Graph Theory we do not require incident edges to have different colors.
\begin{definition}
	Let~$G$ be a graph. A coloring of edges $\delta\colon  E_G\rightarrow \{\text{\blue{}, \red{}}\}$ 
	is called a \emph{NAC-coloring},
	if it is surjective and for every cycle in $G$,
	either all edges have the same color, or
	there are at least 2 edges in each color.
%
\end{definition}

\flexrilog\ contains functionality for computing and showing NAC-col\-or\-ings of a given graph.
The standard output is a textual list but the colorings can be shown in figures as well.
\begin{IOsage}
  C4 = FlexRiGraph([[0,1],[1,2],[2,3],[0,3]])
  C4.NAC_colorings()
  *[NAC-coloring with red edges [[0, 1], [0, 3]] and blue edges [[1, 2], [2, 3]],
  * NAC-coloring with red edges [[0, 1], [1, 2]] and blue edges [[0, 3], [2, 3]],
  * NAC-coloring with red edges [[0, 1], [2, 3]] and blue edges [[0, 3], [1, 2]]]
  C4.show_all_NAC_colorings()
\end{IOsage}
\graphicalOutput[0.6]{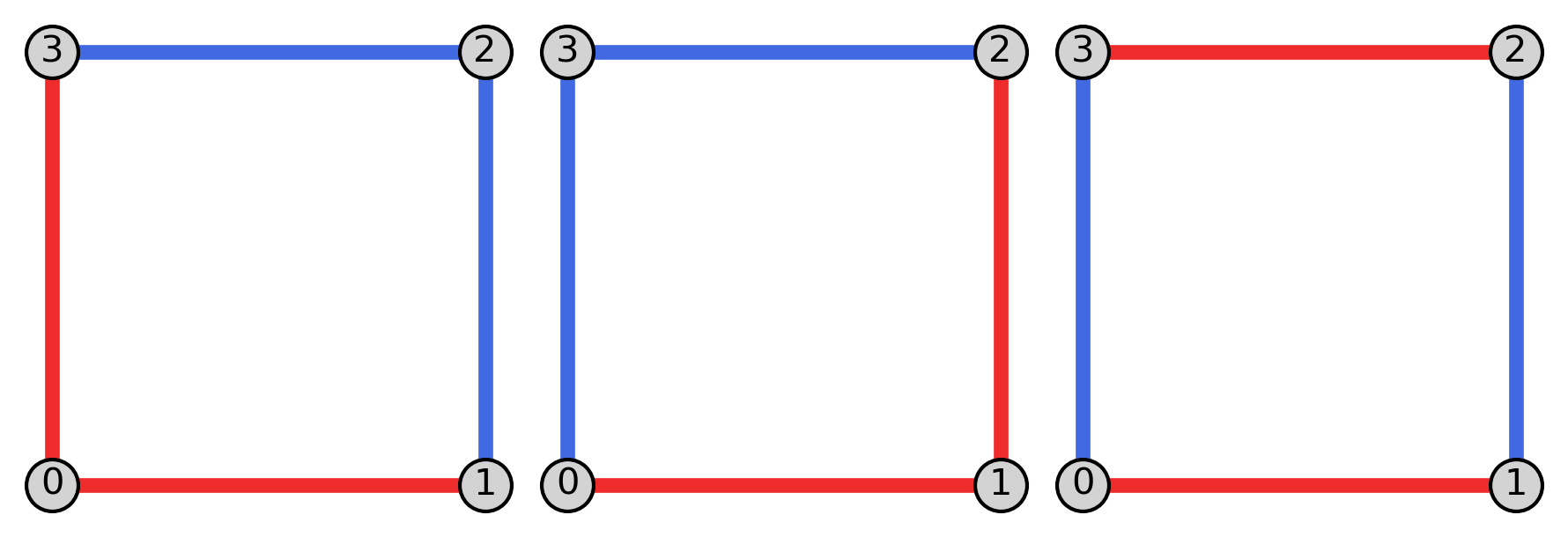}
In order to compute all NAC-colorings, lists of edges that necessarily have the same color
due to being in 3-cycles (so called \emph{$\triangle$-connected components}, see~\cite{flexibleLabelings})
are determined. So far, we then test all possible combinations
how to color them by \red{} and \blue{} \cite[Lemma~2.4]{flexibleLabelings}.
The edges colored the same in the following picture must have the same color in any NAC-coloring,
but no combination satisfies the conditions of NAC-coloring.
\begin{IOsage}
  from flexrilog import GraphGenerator
  N = GraphGenerator.NoNACGraph()
  N.has_NAC_coloring()
  *False
  N.plot(show_triangle_components=True)
\end{IOsage}
\graphicalOutput[0.3]{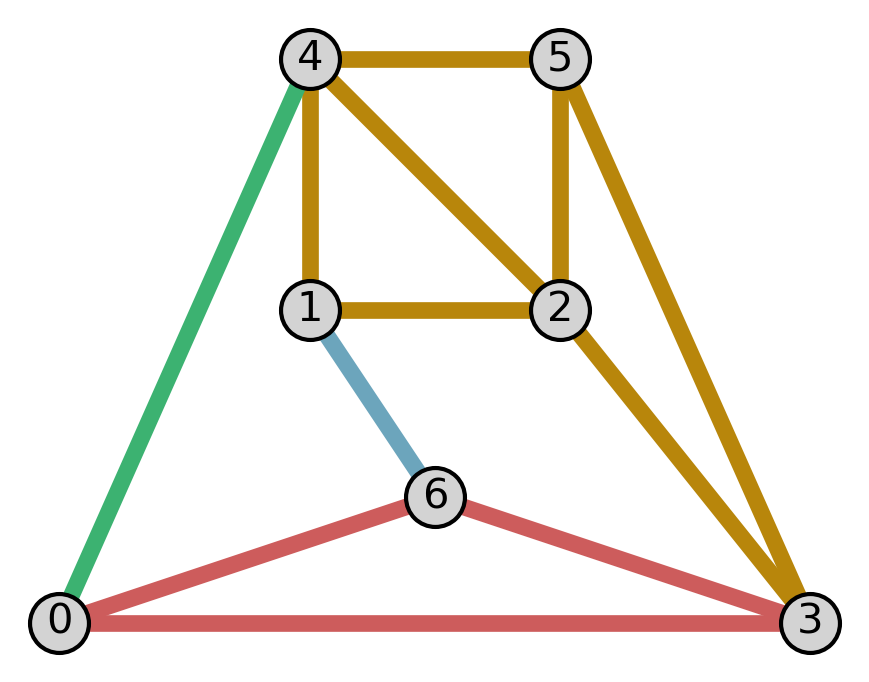}

For graphs with symmetries we get many similar colorings, in a sense that after applying the symmetry one NAC-coloring yields the other.
We call such NAC-colorings isomorphic.
In order to visualize this, NAC-colorings can be named so that isomorphic ones have the same Greek letter but differ by their index.
\begin{IOsage}
  C4.set_NAC_colorings_names()
  C4.NAC_colorings_isomorphism_classes()
  *[[alpha1: NAC-coloring with red edges [[0, 1], [0, 3]] and blue edges [[1, 2], [2, 3]],
  *  alpha2: NAC-coloring with red edges [[0, 1], [1, 2]] and blue edges [[0, 3], [2, 3]]],
  * [beta: NAC-coloring with red edges [[0, 1], [2, 3]] and blue edges [[0, 3], [1, 2]]]]
\end{IOsage}

\subsection{Constructing Motions}\label{sec:motions}
Given a NAC-coloring we are able to construct a motion. The following result from~\cite{flexibleLabelings} describes the relation.
\begin{theorem}\label{thm:nacflexible}
	A connected non-trivial graph has a flexible labeling if and only if it has a NAC-coloring.
\end{theorem}
The main idea to construct a flex is to place the vertices on a grid in such a way that all the edges lie on grid lines.
This can be achieved by placing vertices according to the color component of the graph.
For color components we remove all edges of the other color and then take connected components of the remaining graph.
Then all vertices which lie in the same red component are placed in the same column of the grid and
all vertices from the same blue component are placed in the same row of the grid.
By this procedure each vertex is assigned a unique grid point and all edges of the graph lie on the grid lines.
In \flexrilog\ this can be done with the classmethod \verb=GraphMotion.GridConstruction=.
\begin{IOsage}
  from flexrilog import GraphMotion, GraphGenerator
  P = GraphGenerator.ThreePrismGraph()
  delta = P.NAC_colorings()[0]
  motion_P = GraphMotion.GridConstruction(P, delta)
  motion_P.parametrization()
  *{0: (0, 0),
  * 1: (sin(alpha) + 1, cos(alpha)),
  * 2: (2*sin(alpha) + 1, 2*cos(alpha)),
  * 3: (2*sin(alpha), 2*cos(alpha)),
  * 4: (sin(alpha), cos(alpha)),
  * 5: (1, 0)}
\end{IOsage}
There is also the option to generate an animated SVG showing the NAC-coloring,
which is automatically displayed when used in a Jupyter notebook (the picture below is a screenshot).
If the \verb+fileName+ is specified, the SVG animation is stored and a web browser can be used to view it.
Note that not all web browsers support SVG animations.
It can be chosen, whether the edges are colored according to the NAC-coloring in use.
The package also distinguishes the vertex layout depending on whether it is drawing a graph having no specific placement properties (dark vertices),
or drawing a motion, in which edge lengths are fixed (light vertices).
\begin{Isage}
  motion_P.animation_SVG(edge_partition="NAC",
  .                       fileName="3-prism_grid")
\end{Isage}
\graphicalOutput[0.2]{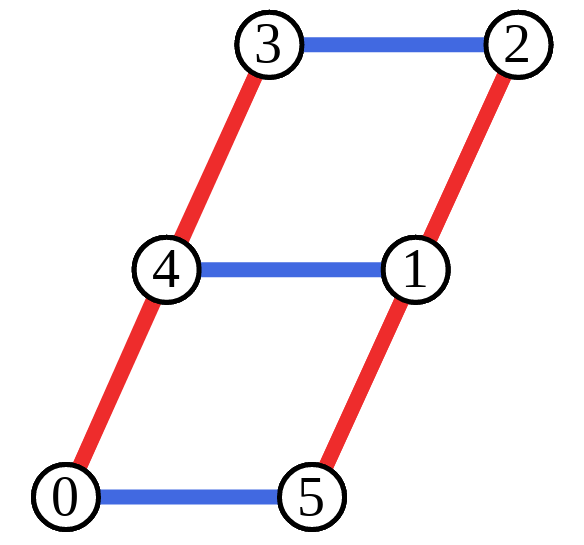}
More generally the base points of the grid can be chosen arbitrarily to get a zig-zag grid.
This can be used to avoid degenerate subgraphs.
Base points consist of two lists. The standard values consists of lists with points $(i,0)$ and $(0,i)$ respectively.
Using them we get a rectangular initial grid.
A zig-zag grid in general does not need to be initially rectangular.
It is uniquely determined by the base points and drawing parallel lines.
Doing so the grid consists of parallelograms.
Usually the grid itself is not easily visible from the output motion.
\begin{Isage}
  motion_P = GraphMotion.GridConstruction(P, delta, 
  .  zigzag=[[[0,0], [3/4,1/2], [2,0]],
  .           [[0,0], [1,0]]])
  motion_P.animation_SVG(edge_partition="NAC")
\end{Isage}
\graphicalOutput[0.2]{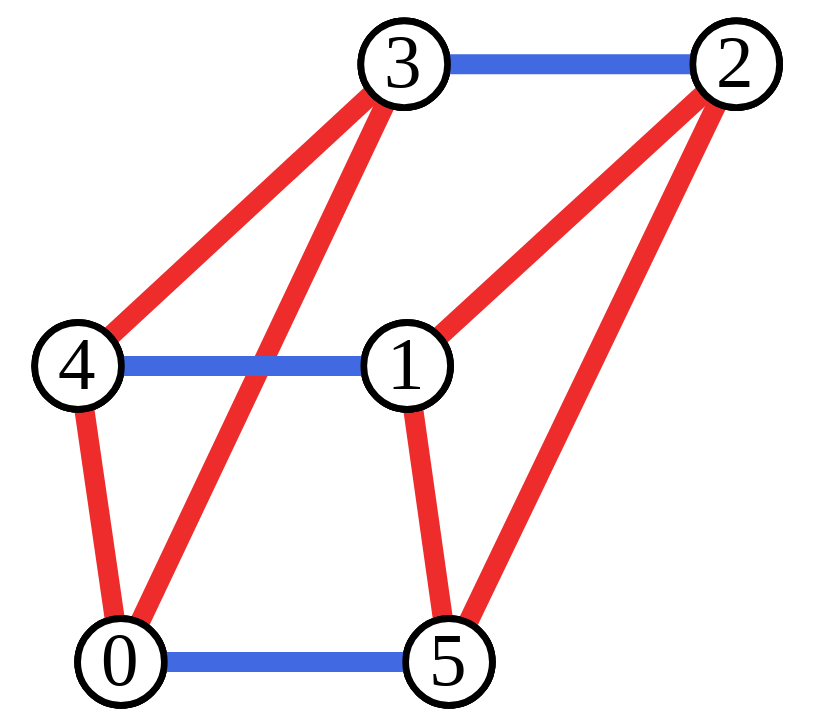}

\section{Movable graphs}\label{sec:movable}
Using the grid construction, non-adjacent vertices might overlap, i.e., the constructed framework is not proper.
Note, that this cannot be avoided by zig-zag constructions either but depends solely on the NAC-coloring in use.
For some graphs all NAC-colorings result in overlapping vertices.
In \flexrilog\ it can be checked whether this is the case.
\begin{IOsage}
  P.has_injective_grid_construction()
  *True
  Q1 = GraphGenerator.Q1Graph()        # see the picture below
  Q1.has_injective_grid_construction()
  *False
\end{IOsage}
For some graphs, a proper flexible labeling exists due to the following lemma~\cite{movableGraphs},
which relates movability to spatial embeddings.
\begin{lemma}
	Let $G$ be a graph with an injective embedding $\omega:V_G\rightarrow\RR^3$ such that for every edge 
	$uv\in E_G$, the vector $\omega(u)-\omega(v)$ is parallel to one of the four vectors 
	$(1,0,0)$, $(0,1,0)$, $(0,0,1)$, $(-1,-1,-1)$, and all four directions are present.
	Then $G$ is movable.
	Moreover, there exist two NAC-colorings of $G$ such that two edges are parallel in the embedding $\omega$
	if and only if they receive the same pair of colors.
\end{lemma}
The package tries to construct such a spatial embedding for all pairs of NAC-colorings.
\begin{IOsage}
  inj, nacs = Q1.has_injective_spatial_embedding(
  .                      certificate=True); inj
  *True
  graphics_array([d.plot() for d in nacs])
\end{IOsage}
\graphicalOutput[0.6]{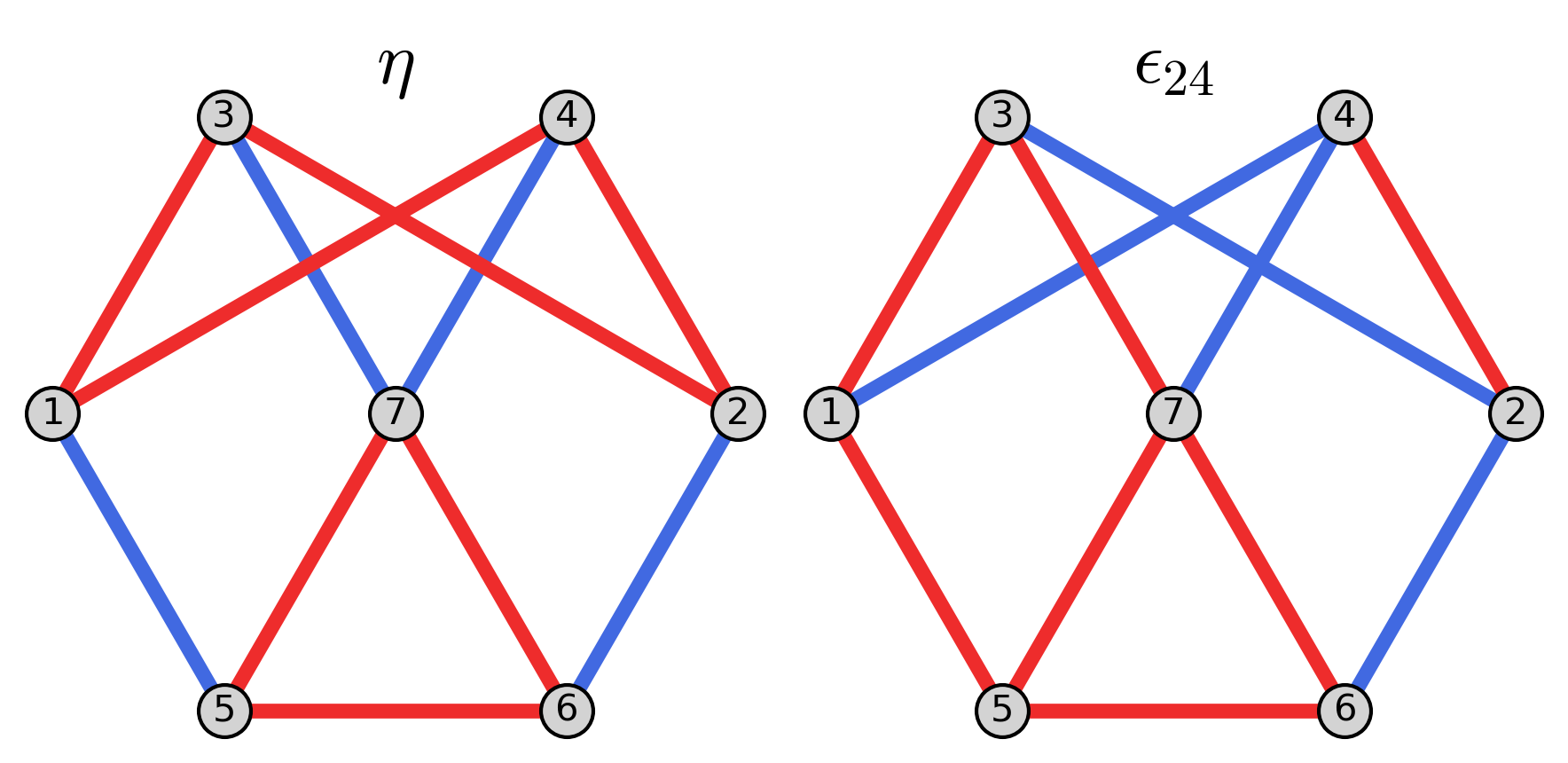}

From the spatial embedding we can construct a motion of the graph.
The motion can be transformed in such a way that a particular edge is fixed.
\begin{IOsage}
  motion_Q1 = GraphMotion.SpatialEmbeddingConstruction(Q1, nacs)
  motion_Q1.fix_edge([5,6])
  motion_Q1.parametrization()
  *{1: ((3*t^2 - 3)/(t^2 + 1), -6*t/(t^2 + 1)),
  * 2: ((t^4 + 23*t^2 + 4)/(t^4 + 5*t^2 + 4), (6*t^3 - 12*t)/(t^4 + 5*t^2 + 4)),
  * 3: ((4*t^2 - 2)/(t^2 + 1), -6*t/(t^2 + 1)),
  * 4: (18*t^2/(t^4 + 5*t^2 + 4), (6*t^3 - 12*t)/(t^4 + 5*t^2 + 4)),
  * 5: (0, 0),
  * 6: (2, 0),
  * 7: (1, 0)}
  motion_Q1.animation_SVG()
\end{IOsage}
\graphicalOutput[0.3]{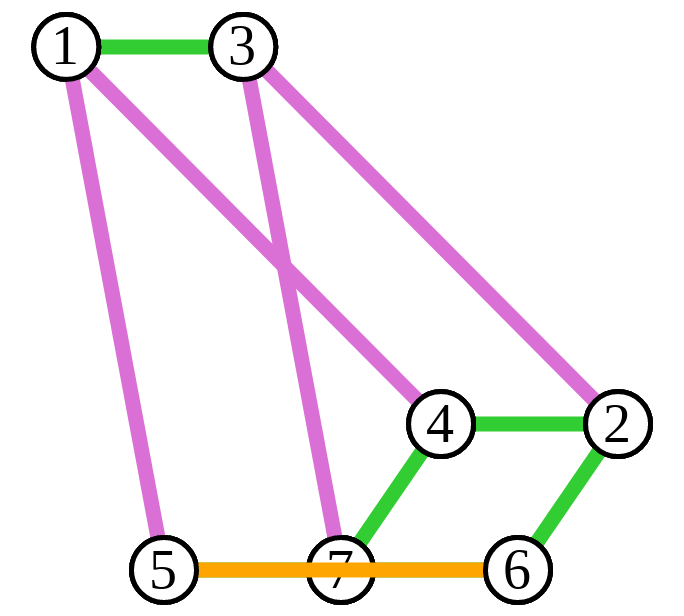}

\noindent Besides the sufficient condition on movability above, there is also a necessary condition given in~\cite{movableGraphs}.
For this condition we consider all NAC-colorings and look for monochromatic paths.
Adding certain edges according to these paths we get a bigger graph with similar movability properties.

For a graph $G$, let $\upairs{G}$ denote the set of all pairs $\{u,v\}\subset V_G$ such that $uv\notin E_G$ and 
there exists a path from $u$ to $v$ which is monochromatic for all NAC-colorings~$\delta$ of $G$.	
If there exists a sequence of graphs $G_0, \dots, G_n$ such that
	$G=G_0$,
	$G_i=(V_{G_{i-1}},E_{G_{i-1}} \cup \upairs{G_{i-1}})$ for $i\in\{1,\dots,n\}$, and
	$\upairs{G_n}=\emptyset$,
then the graph $G_n$ is called \emph{the constant distance closure of $G$}, denoted by $\cdc{G}$.

\begin{theorem}
	A graph $G$ is movable if and only if $\cdc{G}$ is movable.
	Particularly, if $\cdc{G}$ is the complete graph, then $G$ is not movable.
\end{theorem}

We can see that the following graph $G$ is not movable
($G_1$ has no NAC-coloring since $\{3,4\},\{5,6\}\in \upairs{G}$, 
hence, $\upairs{G_1}$ are all non-edges of $G_1$). 
\begin{Isage}
  G = GraphGenerator.MaxEmbeddingsLamanGraph(7)
  G.show_all_NAC_colorings()
\end{Isage}
\graphicalOutput{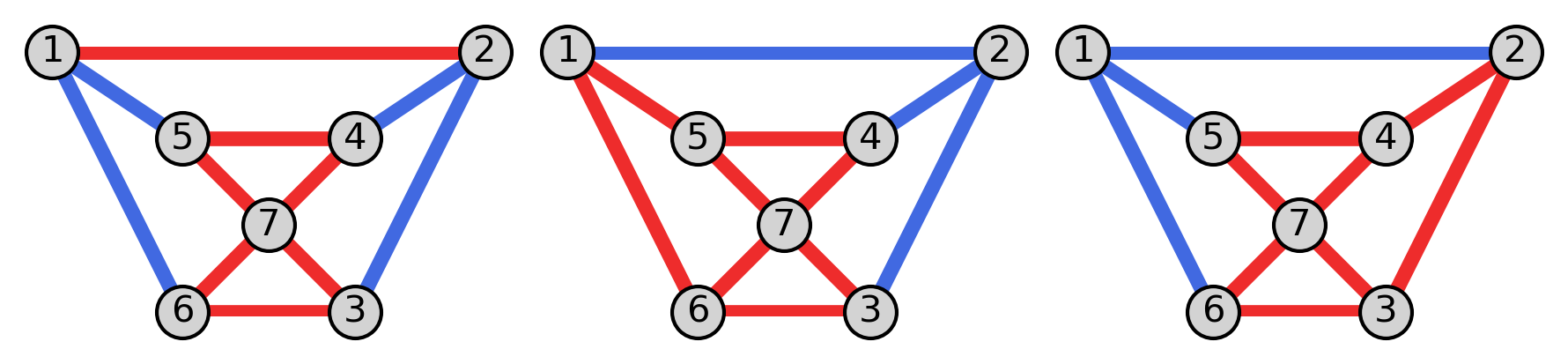}
\vspace{-0.4cm}
\begin{IOsage}
  G.constant_distance_closure().is_complete()
  *True
\end{IOsage}

\section*{Conclusion}
We gave a brief overview of the package and therefore did not cover all functionality.
The package contains a documentation.
As research in the field of flexible and movable graphs is going on the package is further developed,
both regarding improvements as well as new functionality (for instance $n$-fold rotationally symmetric frameworks, see~\cite{RotSymmetry}).
The most current version of \flexrilog\ can be found in \cite{flexrilogGitHub}.

\subsection*{Acknowledgments}
This project was supported by the Austrian Science Fund (FWF): P31061, P31888 and W1214-N15, 
and by the Ministry of Education, Youth and Sports of the Czech Republic, project no. CZ.02.1.01/0.0/0.0/16\_019/0000778.
The project has received funding from the European Union's Horizon~2020 research and innovation programme under the Marie Sk\l{}odowska-Curie grant agreement No~675789.


\begin{thebibliography}{10}

\bibitem{Burmester1893}
L.~Burmester.
\newblock {Die Brennpunktmechanismen}.
\newblock {\em Zeitschrift f{\"u}r Mathematik und Physik}, 38:193--223, 1893.

\bibitem{ZenodoAlg}
J.~Capco, M.~Gallet, G.~Grasegger, C.~Koutschan, N.~Lubbes, and J.~Schicho.
\newblock {An algorithm for computing the number of realizations of a Laman
  graph}.
\newblock Zenodo, 2018.
\newblock \href {https://doi.org/10.5281/zenodo.1245506}
  {\path{doi:10.5281/zenodo.1245506}}.

\bibitem{RotSymmetry}
S.~Dewar, G.~Grasegger, and J.~Legerský.
\newblock Flexible placements of graphs with rotational symmetry, 2020.
\newblock \href {https://arxiv.org/abs/2003.09328} {\path{arXiv:2003.09328}}.

\bibitem{Dixon}
{A. C.} Dixon.
\newblock {On certain deformable frameworks}.
\newblock {\em Messenger}, 29(2):1--21, 1899.

\bibitem{flexrilog}
G.~Grasegger and J.~Legersk{\'y}.
\newblock {FlexRiLoG --- \sage{} package for Flexible and Rigid Labelings of
  Graphs}.
\newblock Zenodo, March 2020.
\newblock \href {https://doi.org/10.5281/zenodo.3078757}
  {\path{doi:10.5281/zenodo.3078757}}.

\bibitem{flexrilogGitHub}
G.~Grasegger and J.~Legersk{\'y}.
\newblock {FlexRiLoG --- \sage{} package for Flexible and Rigid Labelings of
  Graphs, repository}, 2020.
\newblock URL: \url{https://github.com/Legersky/flexrilog/}.

\bibitem{flexibleLabelings}
G.~Grasegger, J.~Legersk{\'y}, and J.~Schicho.
\newblock {Graphs with Flexible Labelings}.
\newblock {\em Discrete {\&} Computational Geometry}, 62(2):461--480, 2019.
\newblock \href {https://doi.org/10.1007/s00454-018-0026-9}
  {\path{doi:10.1007/s00454-018-0026-9}}.

\bibitem{movableGraphs}
G.~Grasegger, J.~Legersk{\'y}, and J.~Schicho.
\newblock {Graphs with Flexible Labelings allowing Injective Realizations}.
\newblock {\em Discrete Mathematics}, in press, 2019.
\newblock \href {https://doi.org/10.1016/j.disc.2019.111713}
  {\path{doi:10.1016/j.disc.2019.111713}}.

\bibitem{ClassificationPaper}
G.~Grasegger, J.~Legerský, and J.~Schicho.
\newblock On the classification of motions of paradoxically movable graphs,
  2020.
\newblock \href {https://arxiv.org/abs/2003.11416} {\path{arXiv:2003.11416}}.

\bibitem{Kempe1877}
{A. B.} Kempe.
\newblock {On Conjugate Four-piece Linkages}.
\newblock {\em Proceedings of the London Mathematical Society},
  s1-9(1):133--149, 11 1877.
\newblock \href {https://doi.org/10.1112/plms/s1-9.1.133}
  {\path{doi:10.1112/plms/s1-9.1.133}}.

\bibitem{Stachel}
H.~Stachel.
\newblock On the flexibility and symmetry of overconstrained mechanisms.
\newblock {\em Philosophical Transactions of the Royal Society of London A:
  Mathematical, Physical and Engineering Sciences}, 372, 2013.
\newblock \href {https://doi.org/10.1098/rsta.2012.0040}
  {\path{doi:10.1098/rsta.2012.0040}}.

\bibitem{sagemath}
{The Sage Developers}.
\newblock {\em {S}ageMath, the {S}age {M}athematics {S}oftware {S}ystem
  ({V}ersion 9.0)}, 2020.
\newblock URL: \url{https://www.sagemath.org}.

\bibitem{WalterHusty}
D.~Walter and {M. L.} Husty.
\newblock On a nine-bar linkage, its possible configurations and conditions for
  paradoxical mobility.
\newblock In {\em 12th World Congress on Mechanism and Machine Science,
  IFToMM}, 2007.

\bibitem{Wunderlich1954}
W.~Wunderlich.
\newblock {Ein merkw{\"u}rdiges Zw{\"o}lfstabgetriebe}.
\newblock {\em {{\"O}sterreichisches Ingenieur-Archiv}}, 8:224--228, 1954.

\bibitem{Wunderlich1976}
W.~Wunderlich.
\newblock On deformable nine-bar linkages with six triple joints.
\newblock {\em Indagationes Mathematicae (Proceedings)}, 79(3):257--262, 1976.
\newblock \href {https://doi.org/10.1016/1385-7258(76)90052-4}
  {\path{doi:10.1016/1385-7258(76)90052-4}}.

\bibitem{Wunderlich1981}
W.~Wunderlich.
\newblock Mechanisms related to {P}oncelet's closure theorem.
\newblock {\em Mechanisms and Machine Theory}, 16:611--620, 1981.
\newblock \href {https://doi.org/10.1016/0094-114X(81)90067-7}
  {\path{doi:10.1016/0094-114X(81)90067-7}}.

\end{thebibliography}

\end{document}